\documentclass[11pt, twocolumn]{article}

\usepackage[left=1.65cm,%
                right=1.65cm,%
                top=2cm,%
                bottom=2cm,]{geometry}%
                
\usepackage[T1]{fontenc}
\usepackage{palatino}
\usepackage{graphicx}
\usepackage[normalem]{ulem}
\usepackage{adjustbox}
\usepackage{booktabs}
\usepackage{makecell}
\usepackage{multirow}

\usepackage{soul}
\usepackage{upgreek}
\usepackage{wrapfig}
\usepackage[switch]{lineno} 

\usepackage{color}
\usepackage[dvipsnames]{xcolor}
\usepackage{hyperref}
\hypersetup{
	colorlinks,
	linkcolor={red!100!black},
	citecolor={blue!100!black},
	urlcolor={blue!80!black}
}
\usepackage{amssymb, amsmath, bm}
\usepackage[tracking]{microtype}
\usepackage{nomencl}

\let\oldequation\equation
\let\oldendequation\endequation
\renewenvironment{equation}
  {\linenomathNonumbers\oldequation}
  {\oldendequation\endlinenomath}
  
\usepackage[labelfont=bf, font={footnotesize}]{caption}
\usepackage{cleveref}
\usepackage{mathrsfs}

\usepackage[indent=00pt]{parskip}

\usepackage{titlesec}

\titleformat*{\section}{\Large\bfseries}
\titleformat*{\subsection}{\large\bfseries}
\titleformat*{\subsubsection}{\normalsize\bfseries}

\titlespacing\section{0pt}{6pt plus 4pt minus 2pt}{4pt plus 2pt minus 2pt}
\titlespacing\subsection{0pt}{6pt plus 4pt minus 2pt}{2pt plus 2pt minus 2pt}
\titlespacing\subsubsection{0pt}{6pt plus 4pt minus 2pt}{2pt plus 2pt minus 2pt}

\usepackage[style=ieee, citestyle=numeric-comp, backend=biber]{biblatex}
\title{
Spider web-inspired sensing and computation with fiber network physical reservoirs 
} 

\author
{
Apoorva Khairnar$^{1}$,
Yogesh Phalak$^{1}$, 
Jun Wang$^{1}$,  
Ziyang Zhou$^{1}$, \\
Benjamin Jantzen$^{2,3}$,
Suyi Li$^{1}$,
Noel Naughton$^{1,^\ast}$
\\[6pt]
\normalsize{\textsuperscript{1}Department of Mechanical Engineering, Virginia Tech, Blacksburg, VA,} \\
\normalsize{\textsuperscript{2}Department of Philosophy, Virginia Tech, Blacksburg, VA,} \\
\normalsize{\textsuperscript{3}Department of Computer Science, Virginia Tech, Blacksburg, VA,} \\
\normalsize{$^\ast$To whom correspondence should be addressed; E-mail: nnaughton@vt.edu}}

\date{}

\begin{document} 

{
\makeatletter
\addtocounter{footnote}{1} 
\renewcommand\thefootnote{\@fnsymbol\c@footnote}%
\makeatother
\maketitle
}

\maketitle 

\begin{abstract}
Physical reservoir computing leverages the intrinsic dynamics of mechanical systems to perform computation through their natural responses to input signals. Here, we study a compliant fiber network inspired by orb-weaving spider webs and investigate how its mechanical design and operating conditions shape its computational capability. Using Cosserat rod-based simulations, we identify how network topology, geometry, actuation, and axial tension impact the nonlinear computation and memory capacity of the network. We further evaluate several readout reduction strategies to assess how computational performance varies with the number and placement of measured outputs.
We then experimentally validate these results using a physical fiber-network prototype. Overall, results provide insights and guidance on design, actuation, and sensing choices to enable fiber networks for mechano-intelligent computation. They demonstrate the ability of structured compliant fibers networks to serve as physical reservoirs capable of nonlinear transformation and input-history retention.

\textbf{Keywords: Physical reservoir computing, Mechanical intelligence, Bio-inspiration, Fiber network}
\end{abstract}

\section{Introduction}

Spider webs, in particular the well-structured webs built by orb-weaving spiders, consist of networks of slender, flexible fibers. 
These webs serve a function critical to the spider's survival, allowing the spider both to capture prey and sense its local environment.
Spiders have evolved a unique ability to interface with and extract information about their environment by sensing vibrations that travel along the web via hairs located on their legs that are highly sensitive to local fiber vibrations. 
Interfacing with the mechanical information embedded in the web enables spiders to localize prey location, identify structural damage, and distinguish between different types of web deformation (e.g., caught prey vs external environmental disturbance)  \cite{wu2023spider, landolfa1996vibrations, mortimer2019spider, masters1984vibrations}.

In this regard, the web serves as an information processing and message passing substrate, mechanically transforming input deformations as a form of mechano-computation  \cite{wang2023mechanical}, which the spider uses to offload complex sensing tasks. 
Inspired by such webs, we consider the information processing capability of an engineered network of compliant fibers through the lens of physical reservoir computing. 

Physical reservoir computing leverages the nonlinear dynamic properties of physical systems to perform complex computations  \cite{tanaka2019recent}. 
Compliant mechanical systems in particular have attracted significant interest due to the ability of their nonlinear dynamics to be leveraged to exhibit 'mechanically intelligent' behavior  \cite{bhovad2021physical, liu2023discriminative, wang2023building, nakajima2013soft, nakajima2014shortterm, nakajima2018soft, degrave2015gait,Khairnar.2025}. By offloading taxing computational tasks to the nonlinear physical dynamics, such systems can simplify sensing and control tasks  \cite{ulrich1988grasping, pfeifer2006body}.

Reservoir computing was originally inspired by recurrent neural structures in the brain and focused on how the nonlinear dynamics of recurrent neural networks with fixed internal weights serve to separate input streams into a high-dimensional state space from which arbitrary functions can be learned via a simple linear output layer of weights  \cite{jaeger2001echo, maass2011liquid, schrauwen2007overview, nakajima2021reservoir}. In time, this insight has been generalized to the realization that any nonlinear dynamical system similarly performs this transformation on the input, meaning that any sufficiently complex, nonlinear dynamical system can serve as a reservoir computer. 

Physical reservoir computing has applied this insight to physical systems, using the nonlinear dynamics of complex systems to perform computation \cite{konkoli2018reservoir, nakajima2020physical, tanaka2019recent}. 
This approach has achieved notable recent interest due to its potential to provide a scalable pathway toward embedding computation into material and structural systems, laying the groundwork for new forms of distributed, embodied intelligence.
Notable examples include nonlinear spring-mass networks capable of emulation and pattern generation tasks  \cite{hauser2011foundation, hauser2012feedback, morales2018mass, urbain2017morphological, dion2018mechanical, du2017memristors, yamane2015wave}, architected materials that respond predictably to input stimuli  \cite{paul2004investigation, paul2006morphological, caluwaerts2011body, zhang2022harnessing}, origami-based mechanical systems that extract complex information from its structural dynamics  \cite{bhovad2021physical, liu2023discriminative, wang2023building, wang2025re}, and soft robotic systems that integrate sensing and actuation to process information  \cite{nakajima2013soft, li2012behavior, nakajima2014shortterm, nakajima2018soft, degrave2015gait, wang2025proprioceptive}.

Critically, the performance of a physical reservoir depends strongly on the design and operating conditions of its mechanical substrate. 
Here, through both simulations and experiments, we investigate the effects of key topological, geometric and actuation features of a network of compliant, overlapping fibers inspired by spiderwebs to establish their ability to serve as physical reservoir computers. 
Further, we consider methods to reduce the dimensionality of the fiber network readouts with minimal loss of performance in order to reduce data redundancy and simplify the fiber network's implementation. 

Overall, our findings indicate that networks of flexible fibers inspired by spider webs can effectively serve as physical reservoir computers. This positions such fiber network reservoirs to potentially serve as unique, physically intelligent structures capable of serving as a low-cost and easy to manufacture compliant sensing technology with potential applications in robotics, autonomous systems, and structural monitoring.

\section{Methods}

\subsection{Fiber network setup}

Inspired by the structured webs of orb-weaving spiders (Fig. \ref{fig1}a), two fiber network topologies are considered: crosshatch and polygonal.
Fiber networks consist of slender, compliant fibers arranged in overlapping architectures with intersecting fibers bonded together. 
Crosshatch topologies consist of an $N\times N$ fiber arrangement of horizontal and vertical fibers (Fig. \ref{fig1}b) while polygonal topologies consist of a $N$-sided polygon where fibers connect all non-neighboring vertices.
Each fiber spans the network with one end fixed in place while the other end is pretensioned and attached to a spring to facilitate fiber deformations (Fig.~\ref{fig1}c). 

To actuate the fiber networks, a lateral mechanical deformation input $u(t)$ is applied to the midpoint of a single fiber (green arrows in Figs. \ref{fig1}b,c). 
The input $u(t)$, which in general may be any continuously varying function, consists of a timeseries constructed by fitting a cubic spline to random points sampled at regular frequency from a uniform distribution.
This approach allows generation of a continuous and smooth input that is compatible with the elastic dynamics of the fiber while still ensuring that any structure observed in the output arises from the reservoir itself rather than from correlations in the input  \cite{dambre2012information}. 
The x--y displacements at $m$ readout locations on the fiber network are tracked as the physical reservoir output $\mathbf{x}(t) = \{x_i(t), y_i(t) \ \forall \ i \in 1...m\}$ where $x_i(t), y_i(t) : \mapsto \mathbb{R}$ and $m$ is determined based on the number of fibers in the network (i.e., $\mathbf{x}(t)$ has $2m$ entries). The readout locations are generally selected to be any location where two fibers cross (nodes) as well as midpoints on fibers between any two nodes.

\subsection{Quantifying physical reservoir capacity}
\label{sec:RC_methods}

Reservoir computers are universal filters  \cite{nakajima2021reservoir} such that, for an arbitrary, real-valued nonlinear functional $z[u(t)]$, there exists a mapping, $\mathcal{W} : \mathbf{x}(t) \mapsto z[u(t)]$ where $\mathcal{W}$ is linear in $\mathbf{x}(t)$. An approximation of such a map,  
\begin{equation}
    \hat{z}(t)=\mathbf{W} \cdot \mathbf{x}(t)
\end{equation}
can be identified by estimating the linear weights $\mathbf{W} \in \mathbb{R}^{2m}$ through ridge regression.
Here ridge regression with a train/test split of 75/25 is used in all cases with splits performed chronologically (e.g. no data shuffling) to prevent information leakage from future samples into training.

In practice, the types of functions that can be learned are limited by the dynamics of the particular reservoir. Our goal then is to probe the capability of a fiber network to serve as a physical reservoir computer. In line with our prior work  \cite{Khairnar.2025}, we do so by defining a general capacity metric capable of quantifying the reservoir's ability to accurately learn a linear map $\mathbf{W}$ that provides an approximation $\hat{z}(t) = \mathbf{W} \cdot \mathbf{x}(t)$ of the functional $z[u(t)]$ over the time interval $[a,b]$  \cite{dambre2012information}
\begin{equation}
c[z] = 1- \frac{\int_{t=a}^b (\hat{z}(t) - z[u(t)])^2 dt}{\int_{t=a}^b (z[u(t)] - \bar{z})^2 dt} \label{eq:c}
\end{equation}
where $\bar{z}$ is the mean of $z[u(t)]$ over the time interval $[a,b ]$.
We then define two specific metrics that quantify the computational capability of the reservoir: nonlinear capacity and memory capacity.

Nonlinear capacity measures the reservoir's ability to perform nonlinear computations on the current input $u(t)$ into the network.
Here, we consider the ability of the reservoir to learn Legendre polynomials $P_k(u(t))$ of order $k$, such that $z_k[u(t)] = P_k(u(t))$ for $k\in\{1,...N\}$ where $N=10$. Legendre polynomials provide a complete and orthogonal basis set that spans the interval $[-1, 1]$ and so provide not only a challenging test of a fiber network's nonlinear capability, but also indicate the ability of the reservoir to approximate any smooth function over this interval due to the efficiency with which Legendre polynomials represent such functions.
We additionally define an overall nonlinear capacity of the reservoir as the sum of the individual capacity metrics $C_{nl} = 1/N \sum_{k=1}^N c[z_k]$.

Memory capacity measures the reservoir's ability to generate linear mappings that recall previous inputs. In this case, $z_{\tau}[u(t)] = u(t-\tau)$ where $\tau\in[0,T]$ is a time lag. We consider recall up to one second in the past ($T=1.0$ s).
Finally, we define the overall memory capacity of the reservoir as the integration of the individual capacity metrics over the recall interval $C_{m} = 1/T \int_{\tau=0}^T c[z_{\tau}]$.

\subsection{Simulation experiments}

To investigate the role of different fiber network parameters, we perform numerical simulations of fiber network dynamics. We model fiber networks as assemblies of Cosserat rods, which provide a computationally efficient description of elastic fibers as one-dimensional filaments while still allowing the full range of deformation modes: bending, twisting, stretching, and shearing.
Each fiber is modeled as an individual Cosserat rod with a diameter of 2 mm, Young's modulus of 100 MPa, and a density of 1000 kg/m\textsuperscript{3}. The diameter and Young's modulus were selected to ensure numerical stability in the simulations.
At every crossing, fibers are coupled using a zero-displacement spring--damper boundary condition  \cite{Zhang:2019}, which enforces connection at the intersection points. 
Each fiber is fixed at one end, while the other end is loaded with a constant tensile force $F_t$. 

To actuate the network, we apply a time-dependent external force $F_{ex}(t) = F_{max}u(t)$ at the midpoint of the central horizontal fiber where $u(t)$ is generated via 5 Hz random sampling.
For all simulations, the timestamp sampling uses a fixed seed such that every network is actuated using the same $u(t)$ with only the force magnitude varied between cases.


Each simulation was 100 seconds long and sampled at 250 Hz with recorded measurements consisting of the planar x--y displacements at fiber intersection points and with the midpoints of fiber segments between intersections. These measurements define the physical reservoir state $\mathbf{x}(t) \in \mathbb{R}^{2m}$. The resulting input-output data is then used within the physical reservoir computing procedure described earlier to evaluate the capacity metrics. 
The ridge regularization parameter $\alpha$ is 0.01 for all simulations except for output feature analysis where it is 0.1 as empirical testing found the best generalization agreement (similar train and test scores) at this level for the output feature analysis.

We implement this numerical modeling approach using \textit{Elastica}  \cite{Gazzola:2018,Zhang:2019,Naughton:2021}, an open-source Cosserat-rod framework in Python that has been applied to and validated against a wide set of fibrous-dynamics problems, including animal locomotion and manipulation  \cite{Zhang:2021, Tekinalp:2024}, fibrous structures  \cite{Bhosale:2022,Khairnar.2025}, and soft robotic control  \cite{Naughton:2021, Chang:2023, Shih:2023, Naughton:2025}.

\begin{figure*}[t!]
    \centering
    \includegraphics[width=\textwidth]{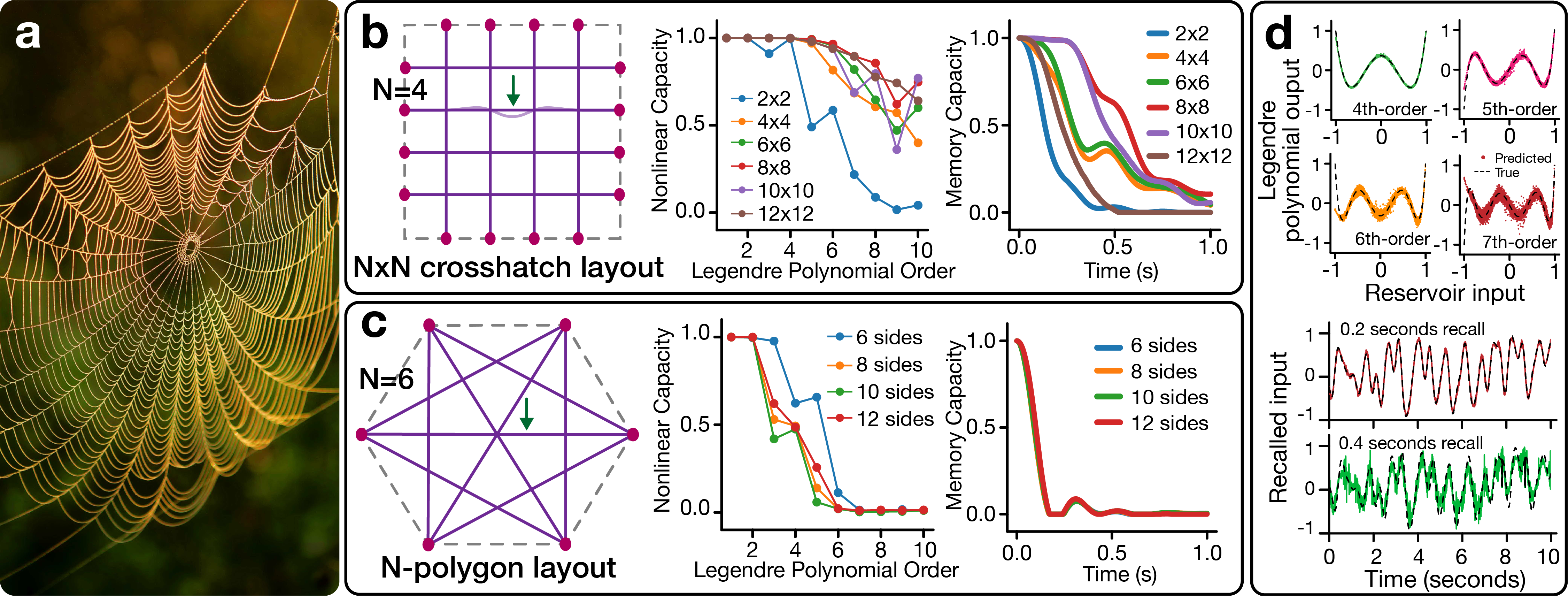}%
    \vspace{-6pt}
    \caption{ \textbf{Spider web-inspired fiber network physical reservoirs.} 
    \textbf{(a)} Spider webs transmit mechanical perturbations from captured prey and the surrounding environment to spiders who process the mechanical information via sensitive hairs on their legs. 
    \textbf{(b)} Cross hatch networks consisting of a $N\times N$ network of horizontal and vertical fibers were examined. Simulation results for an increasing number of $N$ fibers show improvement in both nonlinear and memory capacity, with larger networks generally exhibiting better performance than smaller networks, though occasional performance degradation (e.g. memory capacity of $12\times12$ network) is observed.  
    \textbf{(c)} Polygonal networks, where fibers connect all non-neighboring vertices on a $N$-sided polygon were also considered. Polygons with more vertices, and so more fibers in the network, did not produce improved performance like in the crosshatch networks. The hexagon ($N=6$) network exhibits the best nonlinear capacity while all network topologies had similar memory performance. 
    \textbf{(d)} Example computations of computing nonlinear Legendre polynomials and recalling past inputs for the $8\times8$ crosshatch network, which exhibits the best all-around performance of the network topologies investigated. Black dashed lines denote true values.
    }
    \label{fig1}
    \vspace{-12pt}
\end{figure*}

\subsection{Physical experiments}

Physical experiments were performed for the crosshatch network topology. Fibers consisted of nylon monofilament with lengths of 520 mm and diameters of 0.45 mm. One end of each fiber is fixed, while the other is connected to linear springs ($k=$ 52.5 N/m) to maintain consistent tension (Fig. \ref{fig5}a). Fiber intersections are bonded using a silicone rubber adhesive (Sil-Poxy).

The network was actuated at the midpoint of the central horizontal fiber using a Dynamixel AX-18A servomotor. A lever attached to the servomotor was then attached to the horizontal fiber via a thread, allowing rotation of the servomotor to produce controlled actuation of the fiber network with minimal out-of-plane displacement (Fig. \ref{fig5}b). 
For these experiments, the input $u(t)$ is defined as the normalized vertical displacement of the actuation point rather than the applied force time-series used in simulation since the applied force  at the input location is not directly measured.
The input actuation was generated by modulating the motor joint angle using a cubic spline fitted to random inputs at regularly spaced intervals. The same input profile type was used to modify the motor joint angle as was used to generate the force profile for the simulation experiment (e.g. same random seed), but adjusted such that the profile was generated by fitting a spline to uniform random data sampled at a 2 Hz rate. The resulting spline was then sampled at 100 Hz to generate the input $u(t)$.

Each trial ran for 150 seconds and was recorded at 120 FPS in HD ($1920\times1080$) using a Sony $\alpha$7 III mirrorless camera. 
To measure the reservoir state, red optical tracking markers were attached to fiber intersection points and at the midpoints between intersections, and motion tracked using custom Python code to extract marker trajectories over time.
The x--y displacements of all tracked marker locations were then used to form the reservoir output state $\mathbf{x}(t)$, which was used to compute the same capacity metrics as in the simulation experiments.
For all experiment, the ridge regularization parameter $\alpha$ is fixed at 1 based on empirical testing of generalization performance.

\section{Results}
To identify key fiber architectural features, we first use simulation to explore a wide range of parameter combinations. This allows us to efficiently identify a feature's impact on the mechanical computation and physical reservoir capability of the fiber network setup. 

\begin{figure*}[t!]
    \centering
    \includegraphics[width=\textwidth]{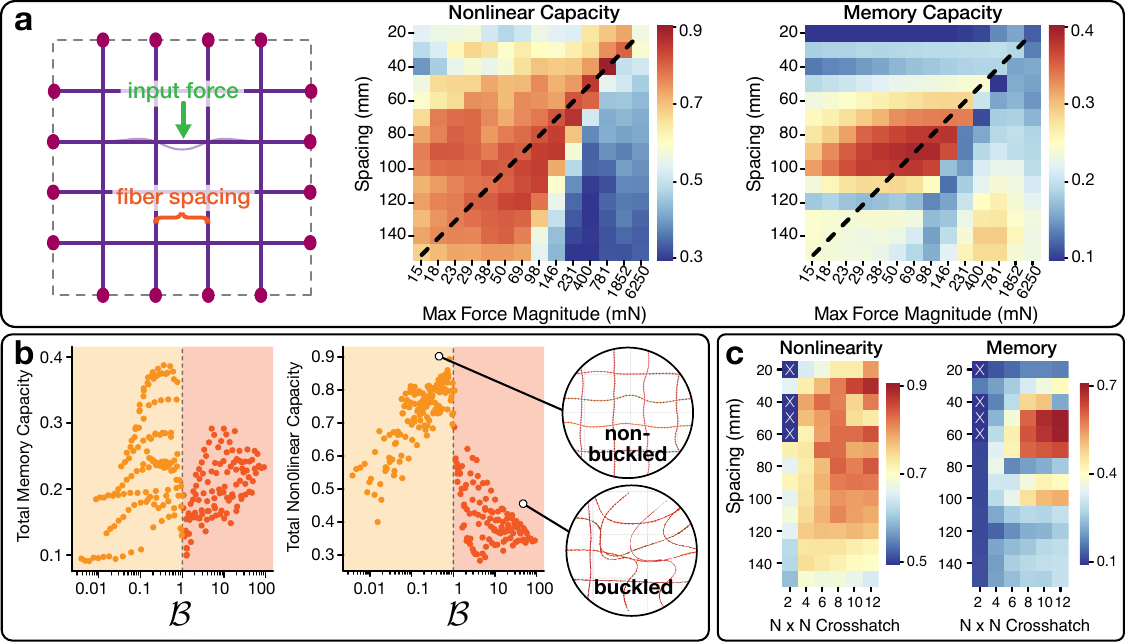}%
    \vspace{-6pt}
    \caption{ \textbf{Crosshatch performance depends on input dynamics matching fiber dynamics.} 
    \textbf{(a)} For a 4x4 crosshatch network, the input force and spacing between fibers was varied. For each fiber spacing, an input force was selected to produce a constant lateral deflection of 13.3 mm for a Euler-Bernoulli simply supported beam (dashed diagonal lines). These input forces were also applied for all other fiber spacings and the total nonlinear capacity (left) and total memory capacity (right) were computed for each case. For each spacing, nonlinear and memory capacity tends to increase for increasing input force before exhibiting a rapid deterioration.
    \textbf{(b)} Plot of the total memory capacity (left) and total nonlinear capacity (right) versus the Euler-Bernoulli buckling criteria where values greater than unity predict buckling onset. The strong performance deterioration in computational capacity observed in (a) is found to be associated with the onset of buckling ($\mathcal{B}>1$) in the vertical fibers.
    \textbf{(c)} Computational performance of networks with increasing number of fibers using the input force associated with the best performance for that fiber spacing in (a). X's indicate numerically unstable simulations that failed to solve. Overall, improved performance is exhibited as the fiber network size increases, with a broad improvement in nonlinear capacity for nearly all fiber spacing while memory performance improvement is more concentrated in fiber spacings between 50 and 70 mm apart.
    }
    \label{fig2}
    \vspace{-12pt}
\end{figure*}

\subsection{Fiber network topology} \label{sec:Topology}

We first consider the role of network topology on reservoir capacity. 
We compare crosshatch and polygonal networks by sweeping $N \in \{2, 4, 6, 8, 10, 12\}$ for the crosshatch case and $N \in \{6, 8, 10, 12\}$ for the polygon case while keeping the nominal network length (outer edge distance) constant. 
As fiber density increases, the spacing between fiber intersection nodes decreases. This effectively stiffens the network, making it more resistant to deformation for the same input force. To ensure consistent levels of deformation across networks, we scale the actuation force to produce a consistent lateral deflection $\delta$ = 13.3 mm based on a model of a simply supported Euler-Bernoulli beam 
\begin{equation}
    F_{max}  = \frac{48 \delta E I}{  s^{3}} \label{eq:force_spacing}
\end{equation}
where $E$ is the fiber Young's modulus, $I$ is the second area moment of inertia, and $s$ is the distance between the two connection nodes of the fiber section where the input force is applied.

For crosshatch networks (Fig. \ref{fig1}b), nonlinear capacity increases as networks become denser up to $8 \times 8$ networks before saturating, with limited to no improvement for the densest networks. Nonlinear capacity improvement is achieved through improved estimation of Legendre polynomials of all orders. In particular, very high capacity (>0.99) is achieved for progressively higher Legendre polynomials as network density increases.
Similar improvement is observed for memory capacity, which rises steadily as fiber density increases up to the $8 \times 8$ network before plateauing for the $10 \times 10$ network and decaying substantially for the $12 \times 12$ network. Larger networks generally support longer recall horizons (up to 0.5 seconds), indicating improved memory of larger delays.
The computational performance of the $8\times 8$ network is illustrated in Fig. \ref{fig1}d.  It shows the performance of the network when challenged to estimate Legendre polynomials of orders four ($c=$ 0.999) through seven ($c=$  0.896)  as well as recall past inputs with time delays of 200 ms ($c=$  0.997) and 400 ms ($c=$  0.713).

In contrast, polygonal networks do not exhibit the same scaling with fiber density (Fig. \ref{fig1}c) indicating that network structure, and not just size, plays an important role in reservoir computational ability.  
All polygonal networks exhibit similar nonlinear and memory performance (though the hexagonal network moderately outperforms the others in nonlinear capacity). Notably, both nonlinear and memory capacity is on par with or below the worst performing crosshatch ($2\times 2$) network.
Together, these results indicate the crosshatch topology consistently outperforms a polygonal topology under the conditions tested. As such, our remaining analyses focus on crosshatch networks.

\subsection{Input force scaling and fiber spacing} \label{sec:force_spacing}

Matching input and reservoir dynamics is well established to be critical to physical reservoir performance. In Fig. \ref{fig1}, we heuristically scaled the input force to produce comparable  lateral deflections at the input location. To investigate how variations in the input force impact reservoir performance, in Fig. \ref{fig2}a we vary input force scaling and fiber spacing in a $4 \times 4$ crosshatch network to identify its impact on overall nonlinear capacity $C_{nl}$ and overall memory capacity $C_{m}$ (the sum of the individual capacities). The spacing between nodes was varied from 20 to 150 mm. For each spacing, a reference actuation force was determined via Eq. \ref{eq:force_spacing} (black dashed diagonal lines) and all spacing-force combinations were then simulated.

For a fixed spacing, increasing actuation force tends to gradually improve both nonlinear and memory capacity up to a point before a strong drop off occurs (Fig. \ref{fig2}a). This drop off forms a ridge in the heat map that is consistent for both nonlinear and memory capacity. Notably, the maximum performance does not follow the constant displacement scaling but rather appears clustered right at the edge of the performance ridge. This is broadly true for nonlinear capacity while memory capacity appears to also be bounded by fiber spacing, with a strong peak in performance occurring for intermediate spacings of 60 to 100 mm. 

It has been commonly observed that reservoir performance is often maximal when the reservoir operates at the `edge of chaos'  \cite{legenstein2007edge,boedecker2012information}. Our results appear to be consistent with this, as the best reservoir performance is found right before a strong drop beyond which performance is permanently decayed. To identify the source of this performance drop, we consider the role of fiber buckling. Specifically, we consider buckling in the two vertical fibers on either side of the horizontal fiber where the input force is applied. In this context, the fibers can be viewed as columns, each subjected to a maximum compressive load $P = F_{max}/2$ that splits the maximum actuation force. Based on this, we define a non-dimensional buckling number  $\mathcal{B}$ that describes the onset of buckling in the fiber ($\mathcal{B} \ge1$)
\begin{equation}
    \mathcal{B} = \frac{Ps^2}{\pi^2EI} {.}  \label{eq:F_cr}
\end{equation}
In Fig. \ref{fig2}b, the overall capacity results of Fig. \ref{fig2}a are replotted against this buckling number, showing the onset of the performance ridge seen in the heatmaps of Fig. \ref{fig2}a corresponds with the onset of buckling  $\mathcal{B}=1$ in these vertical fibers. The existence of buckling is further visually confirmed in the renderings of fiber dynamics, clearly showing visibly buckled configurations once this threshold is crossed.

We briefly note that, post-buckling, there is a region in the memory capacity heatmap of Fig. \ref{fig2}a for large fiber spacing and large force magnitudes where memory performance appears to recover. This suggests there may be a post-buckling region of large fiber spacing where memory capacity can be recovered. However, the uniformly poor nonlinear capacity of post-buckling networks limits the utility of such a regime and so it is not considered in detail here. 
Post-buckling configurations can be highly sensitive to their initial positions but can also propagate their buckling behavior through the network in consistent ways. 
The partial recovery of memory but no recovery of nonlinear computation suggests that information related to these two metrics may be located in these different aspects of the fiber network's dynamics.

\begin{figure}[t!]
    \centering
    \includegraphics[width=\linewidth]{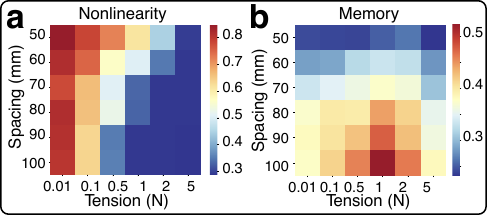}%
    \vspace{-6pt}
    \caption{ \textbf{Fiber pretension and computational capacity.} Total capacity results of a $4\times4$ crosshatch network under different spacing and fiber pretension levels. 
    \textbf{(a)} Total nonlinear capacity declines with pretension, achieving best performance when there is no pretension applied. 
    \textbf{(b)} Total memory capacity increases, generally peaking for 1 N of pretension applied to each fiber before decreasing.  }
    \label{fig3}
    \vspace{-15pt}
\end{figure}

\begin{figure*}[t!]
    \centering
    \includegraphics[width=\textwidth]{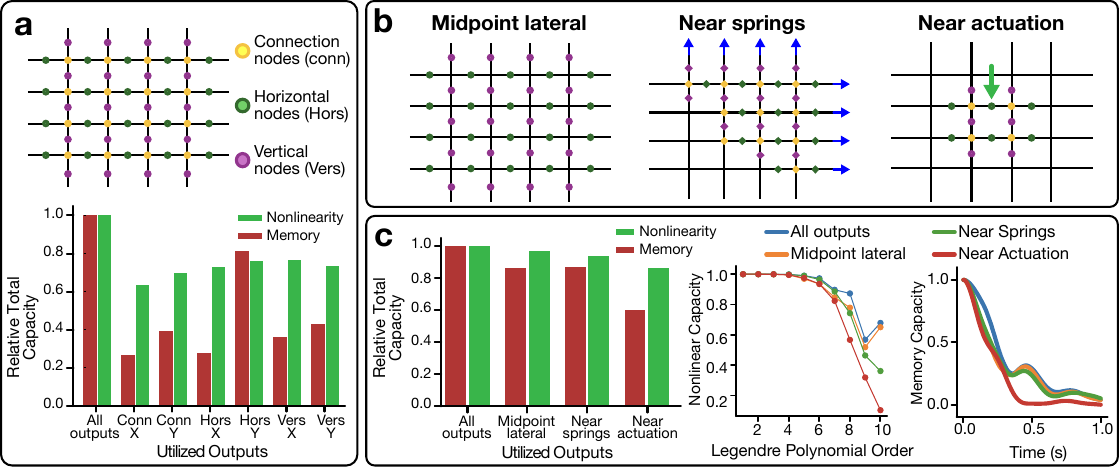}%
    \vspace{-6pt}
    \caption{ \textbf{Readout feature reduction} \textbf{(a)} Readout points are grouped into three groups: fiber intersection points, midpoints on horizontal fibers, and midpoints on vertical fibers. For each group, the x- and y- displacement of the fiber is tracked, yielding 6 total groups. The total nonlinear and memory capacity of each group alone is then computed and compared with the capacity achieved when all outputs are used. 
    \textbf{(b)} Schematics depicting the readout locations of the midpoint lateral, near springs, and near actuation feature groups. For midpoint lateral, only the lateral direction (y-direction on horizontal fibers; x-direction on vertical fibers) is used while near springs and near actuation groups use both x- and y- displacement data for each node in the group. 
    \textbf{(c)} Computational capacity of the three feature sub-groups. Overall, use of midpoint lateral features (Hors-Y and Vers-X) achieves greater than 85\% the performance of all the outputs using only 35\% the number of features.
    }
    \label{fig4}
    \vspace{-12pt}
\end{figure*}

Based on these results, we identified the unbuckled ($\mathcal{B}<1$) force-spacing pairs that yielded the best overall nonlinear capacity and best overall memory capacity and used them to investigate the impact of fiber network size for a fixed spacing between nodes (the results of Fig. \ref{fig1} are for a fixed fiber length with increasingly smaller spacing). $N\times N$ networks of $N \in \{2, 4, 6, 8, 10, 12\}$ were considered (Fig. \ref{fig2}c). 

Generally, both nonlinear capacity and memory capacity increase with network size, though with some notable caveats. While nonlinear capacity increased with number of fibers for spacings less than 120 mm, for spacings greater than this, there is instead a decrease in nonlinear capacity. Improvement in nonlinear capacity is most rapid up to networks of $8\times8$ before plateauing. Further, the improvement in nonlinear capacity is more broad-based than the improvement in memory capacity, with improvement in memory capacity restricted to spacings between 50 and 70 mm, similar to the higher performance band between spacings of 60 and 100 mm identified in Fig. \ref{fig2}a.

\subsection{Fiber pretension}

Prior experimental results have indicated that the axial tension applied to each fiber, $F_t$, can strongly influence memory capacity  \cite{Khairnar.2025}. In our setup, this pretension is applied at one end of every fiber along its axial direction. To understand the impact of this pretension, we investigated the force-spacing pairs that produced the highest memory capacity within the 50-100 mm spacing range and simulated their performance under pretension loads between 0.01 N and 5 N on a $4 \times 4$ network (Fig. \ref{fig3}). 
Nonlinear capacity (Fig. \ref{fig3}a) monotonically decreases as both $F_t$ and fiber spacing increase, with the best nonlinear performance occurring for minimal (0.01 N) pretension. In contrast, memory capacity (Fig. \ref{fig3}b) increases with spacing (for a fixed $F_t$) and also initially increases with  $F_t$ before peaking at 1 N and then decreasing. Pretensioning fibers increases their resonance frequency, potentially suggesting a connection between how memory is stored in the network and fiber dynamics.

\subsection{Output feature down-selection}
To this point, fiber network reservoir performance has been evaluated by using the $(x,y)$ output of fiber crossing points and the midpoints between crossings. For a $N\times N$ network, this yield $6N^2 + 4N$ total features that make up the reservoir readout layer. The use of these features as readout locations has been heuristic, focused on ensuring consistent and uniform coverage of the network.
However, these readout locations exhibit large amounts of co-linearity, indicating redundant information is contained in the reservoir output vector. 
It is natural then to consider if a compact set of readout features can be identified to yield a more optimized reservoir design. 

\begin{figure*}[t!]
    \centering
    \includegraphics[width=\textwidth]{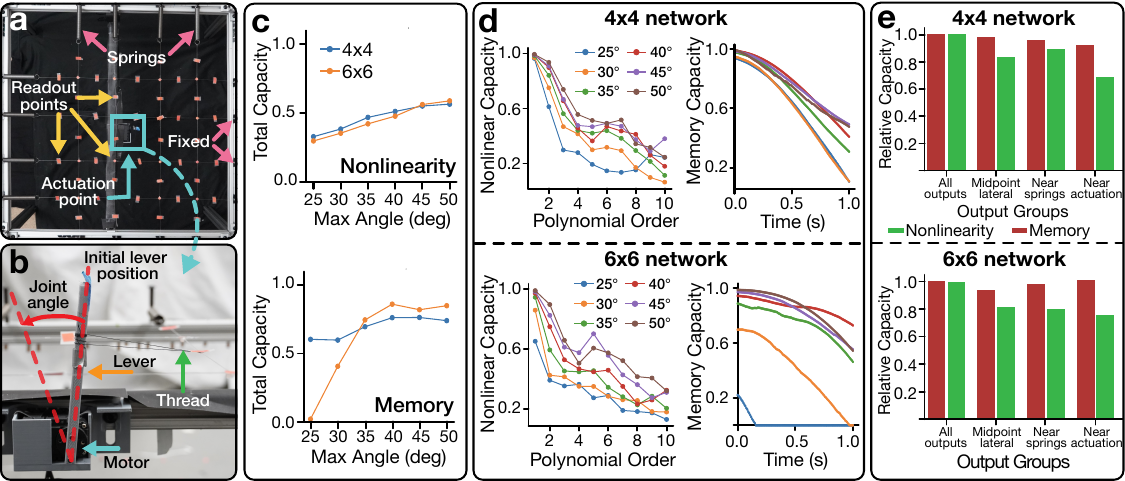}%
    \vspace{-6pt}
    \caption{ \textbf{Physical reservoir fiber network experimental validation.} 
    \textbf{(a)} Overview of experimental setup for $4\times4$ and $6\times6$ crosshatch geometries investigated. 
    \textbf{(b)} Schematic illustrating how the maximum motor angle corresponds to the maximum fiber deflection. 
    \textbf{(c)} Overall nonlinear capacity (Top) and memory capacity (Bottom) experimental results of the $4\times4$ and $6\times6$ network show both networks exhibit similar overall nonlinear capacity, which increases as the maximum deflection angle increases. The $6\times6$ network initially exhibits poor memory capacity for small deflection angles but rapidly improves as the angle increases, eventually outperforming the $4\times4$ network. 
    \textbf{(d)} Individual nonlinear and memory capacity results of $4\times4$ (Top) and $6\times6$ (Bottom) networks for different motor joint angles. 
    \textbf{(e)} Output feature down selection of $4\times4$ (Top) and $6\times6$ (Bottom) networks. Results broadly match simulation results of Fig. \ref{fig4}b. All three groups showed strong retention of memory performance. `Midpoint lateral' and `near springs' groups show similarly strong retention of nonlinear capacity for both networks, with the `near actuation' group showing weaker relative performance.
    }
    \label{fig5}
    \vspace{-12pt}
\end{figure*}

To do so, we adopt an empirical strategy whereby we group the readout points into three groups based on locations at either a fiber crossing point, the midpoint of a horizontal fiber section, or the midpoint of a vertical fiber section. We further consider the x- and y- displacements of each group separately, yielding 6 total subgroups of reservoir outputs. All tests were performed on a $6 \times 6$ crosshatch network with an input force magnitude of 0.85 N, a fiber length of 500 mm, and a tension force of 0.01 N.

In Fig \ref{fig4}a, we examine the nonlinear and memory capacity of each output feature subgroup compared to the computational performance achieved using all of the reservoir outputs. 
Interestingly, fiber connection points exhibited relatively lower computational performance compared to midpoint groups. For nonlinear capacity, performance remained relatively uniform across subgroups. Note that due to co-linearity, this does not indicate that each subgroup has {unique} information, simply that each subgroup contains {some} nonlinear information within it. The y-component of the horizontal midpoint group and the x-component of the vertical midpoint group, which together are the lateral midpoint components, were the two highest performing subgroups. 
For memory capacity, performance was notably highest in the y-component of the horizontal midpoint group with all other subgroups exhibiting substantially lower memory.

Based on these results, we create a feature group consisting of the lateral midpoints (y-component of the horizontal midpoint group and the x-component of the vertical midpoint group) with a total of $2N(N+1)$  features to compare with the full output feature set. We also created two additional \textit{ad hoc} feature groups based on spatial localization of features within the network  (Fig. \ref{fig4}c). The first group, termed `near actuation' consists of readout locations located near the actuation point under the theory that strong dynamics occur in this region of the network. The second group consists of readouts located between the actuation points and the ends of fibers where the pretension force is applied as this region will be most free to deform compared to the opposite side where the fibers are fixed in place. This group is termed as `near springs,' though in these simulations the fiber ends are not attached to physical springs; they are loaded through a constant axial pretension, and the term is used only to indicate proximity to the tensioned boundary.

Of these three feature groups (Fig. \ref{fig4}b), the lateral midpoint and `near springs' group perform best. The lateral midpoint group achieved >95\% the nonlinear computational performance and >85\% the memory performance of when all the output features were used but with a two-third reduction in output features. This result indicate the majority of the mechanical computation of the fiber network is captured by the lateral bending deformations of the fibers.
The `near springs' group performed similarly, with slightly worse nonlinear and slightly better memory performance compared to the lateral midpoint group. 
The near-actuation group performed worse, achieving $\sim$85\% the nonlinear computation performance of when all outputs were used but with $\sim$60\% the memory performance.

\subsection{Experimental validation}
In Fig. \ref{fig5} we seek to confirm our simulation results via physical experiments on $4 \times 4$ and $6 \times 6$ crosshatch networks across a range of input magnitudes for a fixed total fiber length. Motor joint angle was used as a proxy for input force, with the maximum motor angle varied from 25\textdegree{} to 50\textdegree{} in steps of 5\textdegree{} (Fig. \ref{fig5}a).

As actuation magnitude increases, both network sizes show improved nonlinear performance (Fig. \ref{fig5}b), consistent with simulations. Both networks generally exhibit similar overall nonlinear capacity for a given maximum motor joint angle. When considering the specific capacities for different Legendre polynomial orders (Fig. \ref{fig5}c), as maximum joint angle increases, the networks become progressively better at approximating individual Legendre polynomials. 

Notable differences in memory performance are observed between the $4\times 4$ and $6\times 6$ networks.
For the $4\times 4$ network, the overall memory capacity increases modestly with actuation before largely plateauing for larger angles (Fig. \ref{fig5}b). The $6 \times 6$ network displays a different progression. At 25\textdegree{}, its overall memory capacity is close to zero. As actuation increases from 25\textdegree{} to 40\textdegree{}, memory increases sharply before similarly plateauing for higher angles. Notably, for these higher angles, memory capacity is substantially better than the simulated networks, with the $6 \times 6$ network able to  reconstruct inputs from up to 0.5 seconds ago with a capacity above 0.8.
Together, these experimental results confirm key observation from simulations that indicate crosshatch networks of compliant fibers can effectively serve a physical reservoirs capable of both nonlinear computation and memory recall.

\begin{figure}[t!]
    \centering
    \includegraphics[width=\linewidth]{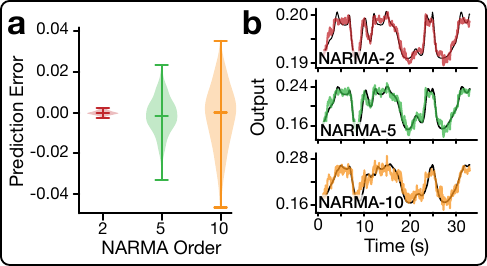}%
    \vspace{-6pt}
    \caption{ \textbf{NARMA-$n$ benchmark performance.} 
    \textbf{(a)} Violin plots of error ($\hat{y}(t) - y(t)$ where $\hat{y}(t)$ is the fiber reservoir prediction of the timeseries at the next timestep) of NARMA-$n$ benchmark tasks for  $n=$ 2, 5, and 10. The fiber reservoir provides unbiased estimation for all tasks, however, its error tails increase for higher values of $n=$, consistent with the increased nonlinearity and memory difficulty of higher order NARMA tasks.
    The root mean square error (RMSE) of the $n=$ 2, 5, and 10 tasks is 0.00096, 0.00924, and 0.01573, respectively, while the interdecile range (IDR) of the error is 0.00246, 0.02383, and 0.04082, respectively.
    \textbf{(b)} Reservoir output estimates of the three NARMA-$n$ timeseries.  }
    \label{fig6}
    \vspace{-15pt}
\end{figure}

We further evaluate if the feature-selection insights from Fig. \ref{fig4} apply to physical networks. We defined the same three output feature groupings as in Fig.~\ref{fig4}b,c and computed the overall nonlinear and memory capacities of the network using only readouts from these subsets (Fig. \ref{fig5}d). 
Results for both network sizes broadly agree with the simulation results of Fig. \ref{fig4}, with the lateral midpoint group and the `near springs' group performing strongly and the `near actuation' group performing worse. Notably, memory performance was retained to a higher degree than nonlinear capacity, a flip of the simulation results of Fig. \ref{fig4}.

Finally, we tested the performance of our physical $6\times 6$ network with a 50\textdegree{} maximum motor joint angle on the NARMA-$n$ benchmark tasks, which have become a common benchmark for evaluating physical reservoir performance  \cite{wringe2025reservoir}.
Experimental trajectories were down-sampled from 120 Hz to 10 Hz and the input signal $u(t)$ was normalized to lie between 0 and 0.2. NARMA-$n$ task trajectories for $n=$2, 5, and 10 were then generated using the standard benchmark coefficients and update rules suggested by Wringe and colleagues  \cite{wringe2025reservoir}.

Figure \ref{fig6}a shows violin plots of the test set error distribution for the three different NARMA-$n$ tasks while Fig. \ref{fig6}b shows a direct comparison between the benchmark timeseries and the reservoir estimate over the test set. Both plots indicate strong, unbiased estimation performance by the fiber reservoir on all three NARMA-$n$ benchmarks, with error bands increasing for NARMA-5 and NARMA-10, reflective of its stronger nonlinear and memory processing requirements.

\section{Discussion}

This work explores the mechano-computational power of a spider web-inspired physical reservoir. By quantifying physical reservoir capability through the joint metrics of nonlinear and memory capacity, we identified how network topology, actuation magnitude, fiber spacing, network size, and axial pretension impact the ability of the fiber network to serve as a physically computing substrate. 

Overall, the fiber network was sensitive to changes in its input and structure. In particular, the crosshatch fiber topology strongly outperformed a polygonal topology, despite the polygonal topology being more visually similar to a spider web. The exact reasons for the crosshatch's better performance are not clear. It is possible the crosshatch's highly uniform organization may help information propagate further through the network before being damped out, while the polygonal topology's central connection node may damp out information due to the multiple fibers acting to stiffen the mechanical response of that node. Further, the fiber network topologies explored here are symmetric, likely causing redundancy in the mechanical response of different parts of the network. Incorporation of random fiber organization and optimization of the fiber's topological motif may enable further improvement in computational performance. 

Reservoir performance was also found to be maximal immediately before the onset of buckling. Buckling injects a level of chaos into the fiber reservoir dynamics due to its extreme sensitivity to small variations in the fiber position at the onset of buckling. As such, it is not surprising that it destroys computational performance in the reservoir. The ability to predict buckling in fiber networks may provide a useful design tool for future reservoirs, as maximizing the portion of the network that is near, but not undergoing, a buckling response may increase the network's computational performance.

The physical experimental implementation of the fiber network broadly agreed with the Cosserat rod simulation results, demonstrating that, indeed, fiber networks can effectively serve as physical reservoir computers. To ensure numerical stability, the Cosserat rod simulations used in this paper used a softer elastic modulus than the nylon monofilament (fishing line) used in the experiments. This discrepancy may explain some of the limited disagreement that is observed as the combination of stiffer fibers and springs used in the physical implementation may effectively be pretensioning the network, in which case the better memory and worse nonlinear performance is consistent with the pretensioning simulation results of Fig. \ref{fig3}.

Finally, we note that the fiber network design used in this work was focused on demonstrating the computational ability of a network made of compliant fibers, which was successfully accomplished. However, further refinements to the network design, such as non-optical reservoir readouts and enabling environmental signal inputs, rather than relying on a dedicated motor, are necessary to advance fiber network physical reservoirs towards a deployable state where they could be used as a sensor. Such refinements will be the focus of future work.

\section{Conclusion}

Inspired by the ability of spiders to use their webs to sense their environment, this work combined simulations and physical experiments to investigate the ability of a network of compliant fibers to serve as a physical reservoir computer. The impact of network topology, geometry, input scaling, and feature outputs were identified and the ability of a fiber network to perform nonlinear transformations in an input signal and recall prior inputs was demonstrated. These basic hallmarks of physical reservoir computing demonstrate the ability of a fiber network to serve as a mechanically intelligent, embedded computational substrate, with potential applications in robotics, autonomous systems, and structural health
monitoring.

\subsection*{Data Availability}
Simulations described in this manuscript used the open source Python-based \textit{Elastica} software available online at \url{https://github.com/GazzolaLab/PyElastica}. 

\subsection*{Acknowledgments}

Funding for this work was provided by NSF EFRI BEGIN OI \#2422340 (B.J., S.L., N.N.) and NSF DCSD \#2328522 (S.L.). Computational support was provided by Virginia Tech's Advanced Research Computing through use of its Tinkercliffs and OWL cluster.

\printbibliography

\end{document}